# Experimental demonstration of magnetic tunnel junction-based computational random-access memory

Yang Lv[1], Brandon R. Zink[1], Robert P. Bloom[1], Hüsrev Cılasun[1], Pravin Khanal[2], Salonik Resch[1], Zamshed Chowdhury[1], Ali Habiboglu[2], Weigang Wang[2], Sachin S. Sapatnekar[1], Ulya Karpuzcu[1] and Jian-Ping Wang[1*]

[1]Department of Electrical and Computer Engineering, University of Minnesota, Minneapolis, Minnesota 55455, USA

[2]Department of Physics, University of Arizona, Tucson, Arizona 85721, USA

*e-mail: jpwang@umn.edu;

# Abstract

Conventional computing paradigm struggles to fulfill the rapidly growing demands from emerging applications, especially those for machine intelligence, because much of the power and energy is consumed by constant data transfers between logic and memory modules. A new paradigm, called "computational random-access memory (CRAM)" has emerged to address this fundamental limitation. CRAM performs logic operations directly using the memory cells themselves, without having the data ever leave the memory. The energy and performance benefits of CRAM for both conventional and emerging applications have been well established by prior numerical studies. However, there lacks an experimental demonstration and study of CRAM to evaluate its computation accuracy, which is a realistic and application-critical metrics for its technological feasibility and competitiveness. In this work, a CRAM array based on magnetic tunnel junctions (MTJs) is experimentally demonstrated. First, basic memory operations as well as 2-, 3-, and 5-input logic operations are studied. Then, a 1-bit full adder with two different designs is demonstrated. Based on the experimental results, a suite of modeling has been developed to characterize the accuracy of CRAM computation. Scalar addition, multiplication, and matrix multiplication, which are essential building blocks for many conventional and machine intelligence applications, are evaluated and show promising accuracy performance. With the confirmation of MTJ-based CRAM's accuracy, there is a strong case that this technology will have a significant impact on power- and energy-demanding applications of machine intelligence.

# Introduction

Recent advances in machine intelligence[1,2] for tasks such as recommender systems[3], speech recognition[4], natural language processing[5], and computer vision[6], have been placing growing demands on our computing systems, especially for implementations with artificial neural networks. A variety of platforms are used, from general-purpose CPUs and GPUs[7,8], to FPGAs[9], to custom-designed accelerators and processors[10–13], to mixed- or fully- analog circuits[14–20]. Most are based on the Von Neumann architecture, with separate logic and memory systems. As shown in Fig. 1a, the inherent segregation of logic and memory requires large amounts of data to be transferred between these modules. In data-intensive scenarios, this transfer becomes a major bottleneck in terms of performance, energy consumption, and cost[21–23]. For example, the data movement consumes about 200 times of the energy used for computation when reading three 64-bit source operands from and writing one 64-bit destination operand to an off-chip main memory[21]. This bottleneck has long been studied. Research aiming at connecting logic and memory more closely has led to new computation paradigms.

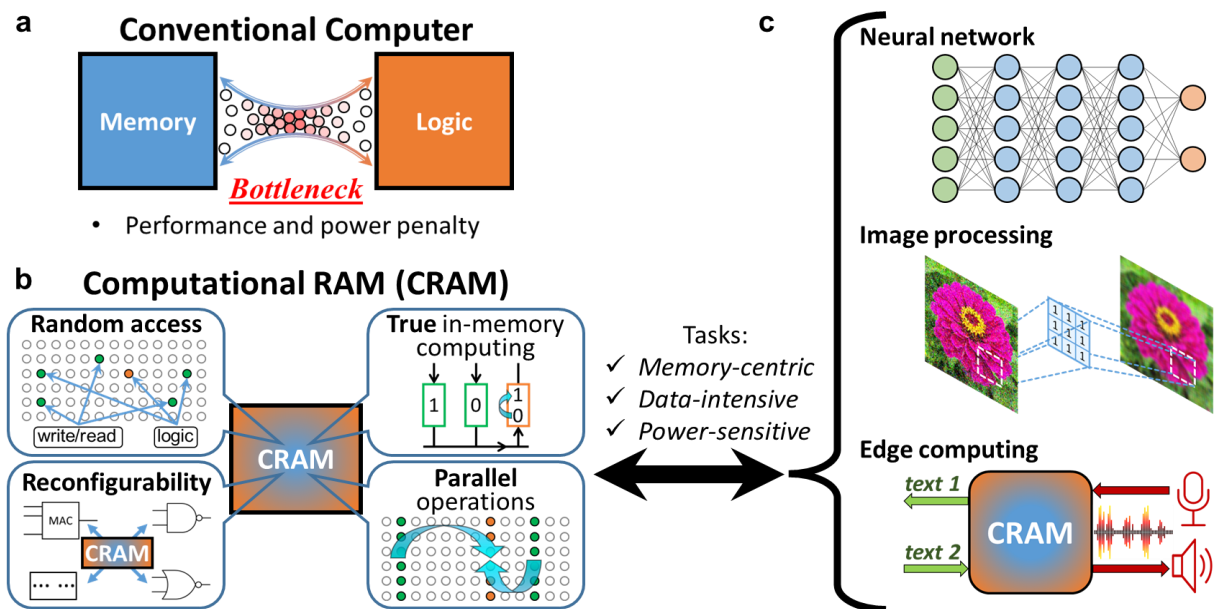

**Fig. 1 | Illustrations of CRAM concept, features, and potential applications. a, b** Compared to a conventional computer architecture (**a**), which suffers from the memory-logic transfer bottleneck, CRAM (**b**) offers significant power and performance improvements. Its unique architecture allows for computation in memory, as well as, random access, reconfigurability, and parallel operation capability. **c** The CRAM could excel in (**c**), data-intensive, memory-centric, or power-sensitive applications, such as neural networks, image processing, or edge computing.

Promising paradigms include "near-memory" and "in-memory" computing. Near-memory processing brings logic physically closer to memory by employing 3D-stacked architectures[24–29]. In-memory computing scatters clusters of logic throughout or around the memory banks on a single chip[14-20,30–35]. Yet another approach is to build systems where the memory itself can perform computation. This has been dubbed "true" in-memory computing[36–42]. The computational random-access memory (CRAM)[38,40] is one of the true in-memory computing paradigms. Logic is performed natively by the memory cells; the data for logic operations never has to leave the memory (Fig. 1b). Additionally, CRAM operates in a fully digital fashion, unlike most other reported in-memory computing schemes[14–20], which are partially or mostly analog. CRAM promises superior energy efficiency and processing performance for machine intelligence applications. It has unique additional features, such as random-access of data and operands, massive parallel computing capabilities, and reconfigurability of operations[38,40]. Also note that although the transistor-less (crossbar) architecture employed by most of the previous true-in-memory computing paradigms[36,37,39,42] allows for higher density, the maximum allowable size of the memory array is often severely limited due to the sneak path issues. CRAM includes transistors in each of its cells for better controlled electrical accessibility, and therefore, larger array size.

The CRAM was initially proposed based on the MTJ device[38], an emerging memory device that relies on spin electronics[43]. Such "spintronic" devices, along with other non-volatile emerging memory devices, usually referred as "X" for logic applications, have been intensively investigated over the past several decades for emerging memory and computing applications as "beyond-CMOS" and/or "CMOS+X" technologies. They offer vastly improved speed, energy efficiency, area, and cost. An additional feature that is exploited by CRAM is their non-volatility[44]. The MTJ device is the most mature of spintronic devices for embedded memory applications, based on endurance[45], energy efficiency[46], and speed[47]. We note that CRAM can be implemented based on not only spintronics devices, but also other non-volatile emerging memory devices.

In its simplest form, an MTJ consists of a thin tunneling barrier layer sandwiched by two ferromagnetic (FM) layers. When a voltage is applied between the two layers, electrons tunnel through the barrier resulting in a charge current. The resistance of the MTJ is a function of the magnetic state of the two FM layers, due to the tunneling magnetoresistance (TMR) effect[48–50]. An MTJ can be engineered to be magnetically bi-stable. Accordingly, it can store information based on its magnetic state. This information can be retrieved by reading the resistance of the device. The MTJ can be electrically switched from one state to the other with a current, due to the spin-transfer torque (STT) effect[51,52]. In this way, an MTJ can be used as an electrically operated memory device with both read and write functionality. A type of random-access memory, the STT-MRAM [53–56] has been developed commercially, utilizing MTJs as memory cells. A typical STT-MRAM consists of an array of bit cells, each containing one transistor and one MTJ. These are referred to as 1 transistor 1 MTJ (1T1M) cells.

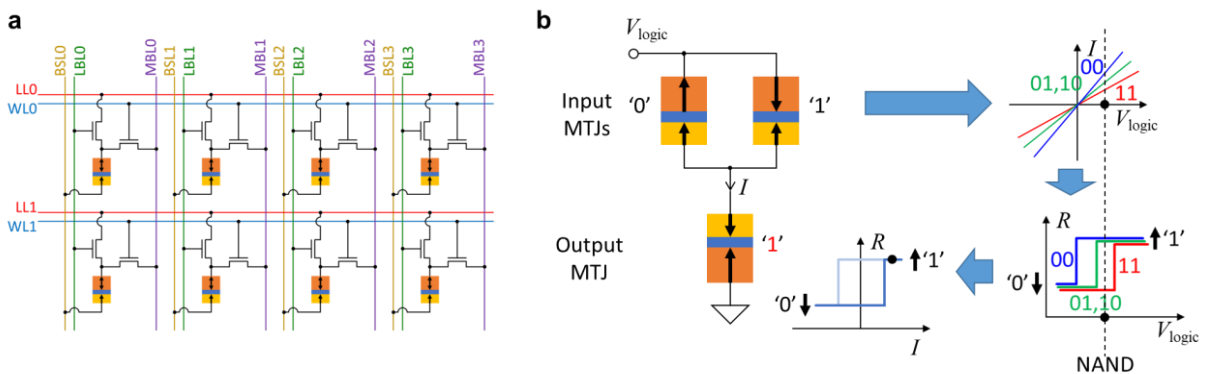

**Fig. 2 | Illustrations of the CRAM cell architecture and the working principle of CRAM logic operation. a** CRAM adopts the so-called 2 transistor 1 MTJ (2T1M) cell architecture. On top of the 1T1M cell architecture of STT-MRAM, additional transistor, as well as the added logic line (LL), and logic bit line (LBL), allow the CRAM to perform logic operations. During a CRAM logic operation, the transistors and lines are manipulated to form equivalent circuit as shown in (**b**). Although CRAM can be built based on various emerging memory devices, we use MTJs and MTJ-based CRAM as an example for illustration purposes. **b** The working principle of CRAM logic operation, the VCL, utilizes the thresholding effect that occurs when switching an MTJ and the TMR effect of MTJ. With an appropriate $V_{logic}$ amplitude, the voltage is translated into different current flowing through the output MTJ by the TMR effect of the input MTJs. Whether the output MTJ switches or not is dependent on the state of input MTJs.

A typical CRAM cell design, as shown in Fig. 2a, is a modification of the 1T1M STT-MRAM architecture[57]. The MTJ, one of the transistors, word line (WL), bit select line (BSL), and memory bit line (MBL), resemble the 1T1M cell architecture of STT-MRAM, which allow the CRAM to perform memory operations. In order to enable logic operation, a second transistor, as well as a logic line (LL), and a logic bit line (LBL), are added to each memory cell. During a logic operation, corresponding transistors and lines are manipulated so that several MTJs in a row are temporarily connected to a shared LL [40]. While the LL is left floating, voltage pulses are applied to the lines connecting to input MTJs with that of the output MTJ being grounded. The logic operation is based on a working principle called voltage-controlled logic (VCL)[58,59], which utilizes the thresholding effect that occurs when switching an MTJ and the TMR effect of MTJ. As shown in Fig. 2b, when a voltage is applied across the input MTJs, the different resistance values result in different current levels. The current flows through the output MTJ, which may or may not switch its state, depending on the states of the input MTJs. In this way, basic bitwise logic operations, such as AND, OR, NAND, NOR, and MAJ, can be realized. A unique feature of VCL is that the logic operation itself does not require the data in the input MTJs to be read out through sense amplifiers at the edge of the array. Rather, it is used locally within the set of MTJs involved in the computation. This is fundamentally why the CRAM computation represents true-in-memory computing: the computation does not require data to travel out of the memory array. It is always processed locally by nearby cells. We note that this concept would also work with other two-terminal stateful passive memory devices, such as memristors. Accordingly, a CRAM could be implemented with such devices. A CRAM could also be implemented with three-terminal stateful devices, such as spin-orbit torque (SOT) devices. This could result in greater energy efficiency and reliability[60]. Although devices with progressive or accumulative switching behavior, such as spintronic domain wall devices[61,62], can be adopted as well, CRAM would otherwise work the best if adopting bi-stable memory devices with strong threshold switching behavior. As an oversimplified speculation, the performance comparison between CRAMs implemented by various emerging memory devices is expected to roughly follow the comparison between these for memory applications, since CRAM utilizes memory devices in similar manners like in memory application. For example, a CRAM implemented based on MTJs should be expected to offer high endurance and high speed. Also, generally, a CRAM logic operation should consume energy comparable to the energy consumption of a memory write operation, for the same emerging memory device operating at the same speed. However, a careful case-by-case analysis is necessary for CRAMs implemented by each emerging memory device technology. Also note that we do not show a specific circuit design of CRAM peripherals because CRAM does not require significant circuit design change in sensing amplifiers or peripherals compared to 1T1M STT-MRAM. And these in the STT-MRAM are already common and mature. Lastly, the true-in-memory computing characteristic of CRAM is limited to within a continuous CRAM array: any computation that requires access to data across separate CRAM arrays will require additional data access and movement. And the size of an array is ultimately limited by parasitic effects of interconnects[63]. However, these limitations are true for all other in-memory computing paradigms. CRAM is not at any disadvantage in this scenario.

On top of the potential performance benefits that the emerging memory devices bring, at circuit and architecture level, CRAM fundamentally provides several benefits (Fig. 1b): (1) the elimination of the costly performance and energy penalties associated with transferring data between logic and memory; (2) random access of data for the inputs and outputs to operations; (3) the reconfigurability of operations, as any of the logic operation AND, OR, NAND, NOR, and MAJ can be programmed; and (4) the performance gain of massive parallelism, as identical operations can be performed in parallel in each row of the CRAM array when data is allocated properly. Based on analysis and benchmarking, CRAM has the potential to deliver significant gains in performance and power efficiency, particularly for data-intensive, memory-centric, or power sensitive applications, such as bioinformatics[40,64,65], image[66] and signal[67] processing, neural networks[66,68], and edge computing[69] (Fig. 1c). For example, a CRAM-based machine-learning inference accelerator was estimated to achieve an improvement on the order of 1000× over a state-of-art solution, in terms of the energy-delay product[70]. Another example shows that CRAM (at 10 nm technology node) consumes 0.47 µJ and 434 ns of energy and time, respectively, to perform a MNIST hand-written digit classifier task. It is 2500× and 1700× less energy and time, respectively, compared to a near-memory processing system at 16 nm technology node[66]. And yet, to date, there have been no experimental studies of CRAM.

In this work, we present the first experimental demonstration of a CRAM array. Although based on a small 1×7 array, it successfully shows complete CRAM array operations. We illustrate computation with a 1-bit full adder. This work provides a proof-of-concept as well as a platform with which to study key aspects of the technology experimentally. We provide detailed projections and guidelines for future CRAM design and development. Specifically, based on the experiment results, models and calculations of CRAM logic operations are developed and verified. The results connect the CRAM gate-level accuracy or error rate to MTJ TMR ratio, logic operation pulse width, and other parameters. Then we evaluate the accuracy of a multi-bit adder, a multiplier, and a matrix multiplication unit, which are essential building blocks for many conventional and machine intelligence applications, including artificial neural networks.

## Experiments

Figure 3 shows the experimental setup, consisting of both hardware and software. The hardware is built with a so-called 'circuit-around-die' approach[71]: semiconductor circuitry is built with commercially-available components around the MTJ dies. This approach offers a more rapid development cycle and flexibility needed for exploratory experimental study on CRAM arrays and potential new MTJ technologies, while the major foundries lack the specific process design kit available for making a CRAM array fully integrated with CMOS. The hardware is a 1×7 CRAM array, with the design of cells taken from the 2T1M CRAM cells[38,40], modified for simplified memory access. Software on a PC controls the operation. It communicates with the hardware with basic commands: 'open/close transistors'; 'apply voltage pulses' to perform write and logic operations'; and 'read MTJ cell resistance'. The software collects real-time measurements of the data associated with CRAM operations for analysis and for visualization. All aspects of the 1×7 CRAM array are functional: memory write, memory read, and logic operations (more details in Methods section, and Supplementary Note S1).

MTJs with perpendicular interfacial anisotropy are used in the CRAM. They exhibit low resistance-area (RA) product and high TMR ratio – approximately 100% – when sized at 100 nm in diameter (more details in Supplementary Note S2).

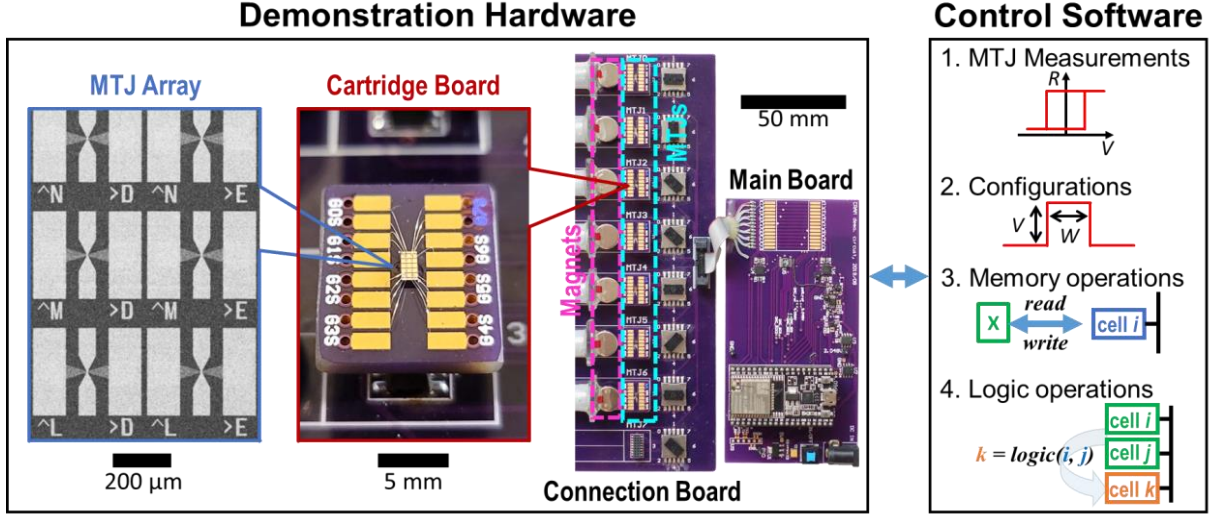

**Fig. 3 | CRAM experimental setup.** The setup consists of custom-built hardware and a suite of control software. It demonstrates a fully functioning 1×7 CRAM array. The hardware consists of a main board hosting all necessary electronics except for the MTJ devices; a connection board on which passive switches, connectors, and magnetic bias field mechanisms are hosted; and multiple cartridge boards that each have a MTJ array mounted and multiple MTJ devices that are wire bonded. The gray-scale scanning electron microscopy image shows the MTJ array used. The color optical photographs show the cartridge board and the entire hardware setup. The software is responsible for real-time measurements of the MTJs; configuration and execution of CRAM operations: memory write, memory read, and logic; and data collection. It is run on a PC, which communicates wirelessly with the main board.

## Results

### I. Device Properties and CRAM Memory Operations

The experiments begin with measuring the resistance (R)–voltage (V) properties of each MTJ device and of each die. In order to compensate for device-to-device variations, the bias magnetic fields for each MTJ are adjusted so that the R–V properties are as close to each other as possible (more details in Supplementary Note S2). As the processes of making CRAM array mature, bias magnetic fields are expected to be no longer needed and all CRAM cells will be able to be operated with uniform parameters and under uniform conditions. The resistance threshold for the MTJs logic states is also determined in this stage.

Then the seven MTJ cells are tested for memory operations with various write pulse amplitudes and widths. Based on the observed write error rates for memory write operations, appropriate pulse amplitudes and widths are configured, achieving reliable memory write operations with an average write error rate of less than $1.5 \times 10^{-4}$ (more details in Supplementary Note S3). We designate logic '0' and '1' to the parallel (P) low resistance state and anti-parallel (AP) high resistance state of MTJ, respectively.

### II. CRAM Logic Operations

Two-input logic operations are studied. The output cell is first initialized by writing '0' to it. Then two input cells are connected to the output cell through the LL by turning on corresponding transistors. Voltage pulses of amplitude of $V_{logic}$, $V_{logic}$, 0, are simultaneously applied to the two-input cells and the output cell, respectively. This is the same as grounding the output cell while applying a voltage pulse of $V_{logic}$ to the two input cells. Then depending on the input cells states, the output cell will have a certain probability of being switched from '0' to '1'. Such a cycle of operations is repeated n times, and the statistical mean of the output logic state, $<D_{out}>$, is obtained. The entire process is repeated for different $V_{logic}$ values and input states. The basis for logic operations in the CRAM is the state-dependent resistance of the input cells. These shift and displace the output cell's switching probability transfer curve. As a result, the output cell switches state based on specific input states, therefore, implementing a logic function such as AND, OR, NAND, NOR, or MAJ. A specific initial state of the output cell and $V_{logic}$ value corresponds to one of these logic gates[66]. The time duration or pulse width of the voltage pulse applied during a logic operation is expected to contribute to most of the time required to complete a logic operation. In the following we use the term logic speed to generally refer to the speed of a logic operation. Logic speed is approximately inversely proportional to the time duration of the voltage pulse used during a logic operation.

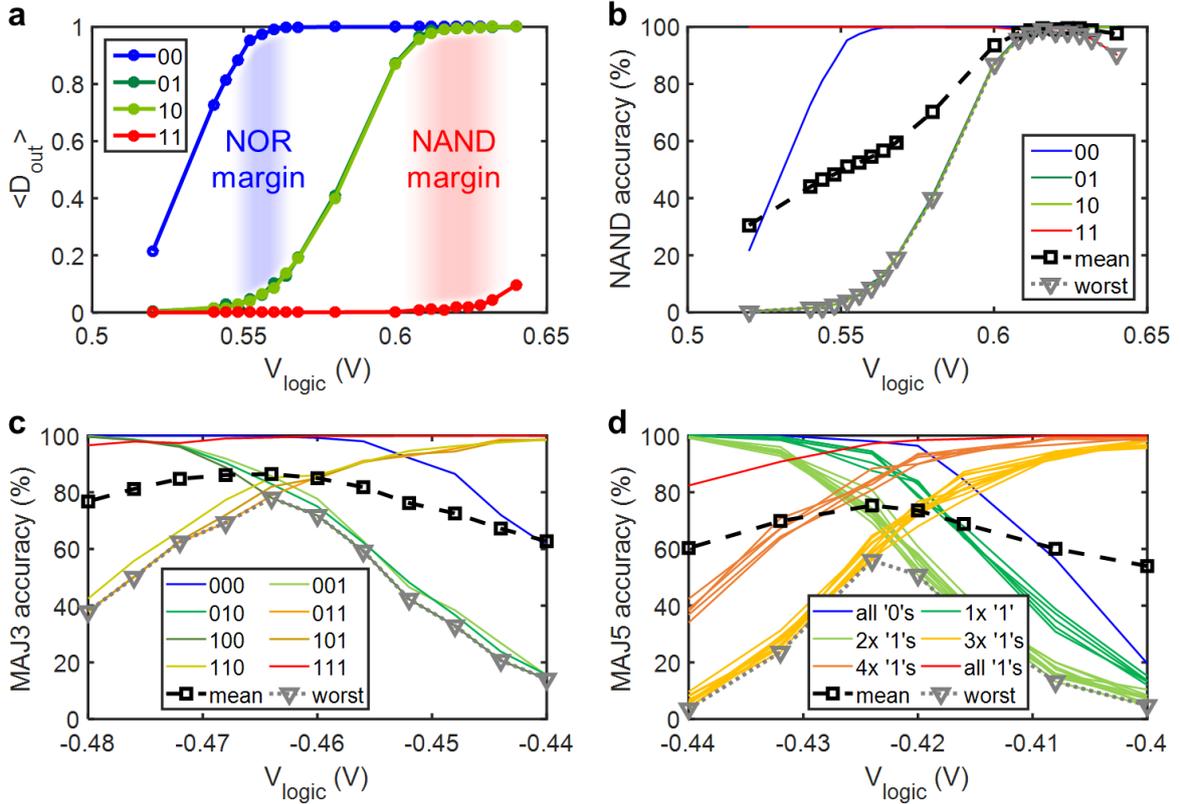

**Fig. 4 | Experimental results for CRAM logic operation. a** Output logic average, $D_{out}$, vs. logic voltage, $V_{logic}$. In a 2-input logic operation, two input cells and one output MTJ cell are involved. The output cell's terminal is grounded, while the common line is left floating. A logic voltage pulse is applied on the two input cells' terminals for a fixed duration (pulse width) of 1 ms. Before each logic operation, input data is written to the input cells. After each logic operation, the output cell's state is read. Each curve corresponds to a specific input state. Each data point represents the statistical average of the output cell's logic state, $<D_{out}>$, sampled by 1000 repeats (n = 1000) of the operations. The separation between the $<D_{out}>$ curves indicates the margins for NOR or NAND operation, highlighted in blue and red, respectively. **b** Accuracy of 2-input NAND operation vs. logic voltage, $V_{logic}$. The results in (**a**) can be converted into a more straightforward metric, accuracy, for the NAND truth table. The curve labeled 'mean', and 'worst' indicate the average and the worst-case accuracy across all input states, respectively. So, for NAND operation, the optimal logic voltage is indicated in such a plot where the mean or worst accuracy is maximized. **C, d** Accuracy of MAJ3 (**c**) and MAJ5 (**d**) logic operation vs. logic voltage, $V_{logic}$. Each curve corresponds to an input state or a group of input states. And each data point represents the statistical average of the output MTJ logic state sampled by n = 1000 and n = 250, for (**c**) and (**d**), respectively.

The experimental results are shown in Figs. 4a and 4b. Generally, for a given input state, $<D_{out}>$ increases with increasing $V_{logic}$. The $<D_{out}>$ response curves are input-state dependent. The four input states can be divided into three groups:
- The '00' input state yields the lowest resistance at the two input cells, so the output cell switches from '0' to '1' first (with the lowest $V_{logic}$).
- The '11' input state yields the highest resistance at the two input cells, so the output cell switches from '0' to '1' last (with the highest $V_{logic}$).
- The '01' and '10' input states both yield resistance that falls in between that of '00' and '11' so that the output cell's response curve falls in between that of '00' and '11'.

Figure 4a shows the experiment results. The two regions highlighted in blue and red that fall in between the three groups of response curves are suitable for NOR and NAND operations, respectively. For example, in the red region, the '11' input has a high probability of yielding a '0' output, while the other three input states have a high probability of yielding a '1' output. This matches the expected truth table for a NAND logic gate. Therefore, if $V_{logic}$ is chosen carefully – within the red region for the CRAM 2-input logic operation – the operation performed is highly likely to be NAND.

The experimental results of $<D_{out}>$ can be converted into a straightforward format representing the accuracy for specified logic function. This translation can be computed by simply subtracting $<D_{out}>$ from 1 for those input states where a '0' output is expected in the truth table of the logic function. Figure 4b shows NAND accuracy of the same 2-input CRAM logic operation. The 'mean' and 'worst' plots are based on the average value and minimum value of the accuracy, respectively, across all input state combinations at a fixed value for $V_{logic}$. Based on the experimental results, if $V_{logic} = 0.624$ or $0.616$ V, the CRAM delivers a NAND operation with a best mean and a worst-case accuracy of about 99.4% and 99.0%, respectively. From a circuit perspective, both increasing the effective TMR ratio of input cells and/or making the output cell's response curve steeper would increase the vertical separation of these input-state-dependent curves, resulting in higher accuracy. For example, higher effective TMR ratio of input cells results in larger contrast of current in the output cell between different input states. Therefore, there is more 'horizontal' room to separate the $<D_{out}>$ curves associated with different input states so that for the inputs with which the output is expected to be '0' or '1', the $<D_{out}>$ of output cell is closer to the expected value ('0' or '1'). Also note that for a logic operation, the 'accuracy' and 'error rate' are essential two quantities describing the same thing: how true is the logic operation is, statistically. By definition, the sum of accuracy and error rate is always 1. The higher or closer to 1 the accuracy is, the better. The lower or closer to 0 the error rate is, the better. Lastly, to facilitate better visualization of how the resistance changes of different input cell states are translated into voltage differences on the output cell resulting in it being switched or unswitched, we list the equivalent resistance of the two input cells combined in parallel and the resulted voltage on the output cell in the following. With $V_{logic} = 0.620$ V, the equivalent resistance of input cells and the resulted voltage on the output cell are 0.4133 V and 1120 Ω, 0.3753 V and 1461 Ω, and 0.3248 V and 2037 Ω, for input states '00', '01' or '10', and '11', respectively. Note that these values are estimated by the experiment-based modeling, which is introduced in the later part of this paper.

With more input cells, we also studied 3-input and 5-input majority logic operations. Figure 4c shows the accuracy of a 3-input MAJ3 logic operation. At $V_{logic} = -0.464$ V, both the optimal mean and the worst-case accuracy are observed to be 86.5% and 78.0%, respectively. Similarly, for a 5-input MAJ5 logic operation, shown in Fig. 4d, both the optimal mean and the worst-case accuracy are observed to be 75% and 56%, respectively. As expected, comparing 2-input, 3-input, and 5-input logic operation, the accuracy decreases with an increasing number of inputs (more discussions and explanations in Supplementary Note S4).

## III. CRAM Full Adder

Having demonstrated fundamental elements of CRAM – memory write operations, memory read operations, and logic operations – we turn to more complex operations. We demonstrate a 1-bit full adder. This device takes two 1-bit operands, A and B, as well as a 1-bit carry-in, C, as inputs, and outputs a 1-bit sum, S, and a 1-bit carry-out, $C_{out}$. A variety of implementations exist. We investigate two common designs: (1) one that uses a combination of majority and inversion logic gates, which we will refer to as a 'MAJ+NOT' design; and (2) one that uses only NAND gates, which we will refer to as an 'all-NAND' design. Figures 5a and 5b illustrate these designs. Supplementary Note S5 provides more details.

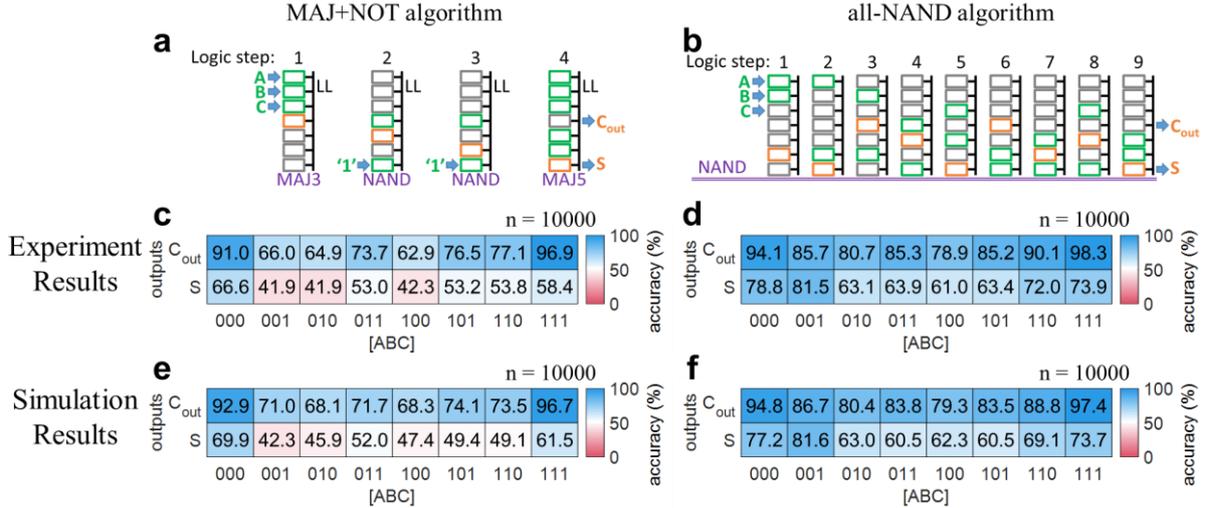

**Fig. 5 | CRAM 1-bit full adder demonstration results. a**, **b** Illustrations of the 'MAJ+NOT' and 'all-NAND' 1-bit full adder designs. Green and orange letter symbols indicate input and output data for the full adder, respectively. From left to right, numbered by 'logic step', each drawing shows the intended input (green rectangle) and output (orange rectangle) cells involved in the logic operation. The text in purple under each drawing indicates the intended function of the logic operation (MAJ3, NAND, or MAJ5). **c**, **d**, **e**, **f** Experimental (**c**, **d**) and simulation (**e**, **f**) results of the output accuracy of 1-bit full adder operations by CRAM with the MAJ+NOT (**c**, **e**) and all-NAND (**d**, **f**) designs. The CRAM adder's outputs, S and $C_{out}$, are assessed against the expected values, i.e., their truth table, for all input states of A, B, and C. The accuracy of each result for each input state is shown by the numerical value in black font, as well as, represented by the color of the box with red (or blue) indicating wrong (or correct), or accuracy of 0% (100%). The accuracy is calculated based on the statistical average of outputs obtained by repeating the full adder execution n times, for n = 10000. The experimental results for the MAJ+NOT (**c**) and all-NAND (**d**) designs are obtained by repeatedly executing the operation for all input states and observing the output states. The simulation results for the MAJ+NOT (**e**) and all-NAND (**f**) designs are obtained with probabilistic modeling, using Monte Carlo methods. The accuracy of individual logic operation is set to what was observed experimentally.

Figures 5c-f show both the experimental and simulation results for the MAJ+NOT and the all-NAND designs, respectively. Each plot is a colormap that lists the accuracy of the output bits S and $C_{out}$, with each input state coded as [ABC]. The blue (red) indicates good/desired (bad/undesired) accuracy. In the boxes of colormap, results in saturated blue are the most desirable. The numerical values of accuracy are also labeled accordingly. From the experimental results for the MAJ+NOT design full adder shown in Fig. 5c, we make two observations:

- The accuracy of $C_{out}$ is generally higher than that of S. This is because $C_{out}$ is directly produced by the first MAJ3 operation from inputs A, B, and C, while S is produced after additional logic operations. We also note that since $C_{out}$ is produced earlier than S, it is less impacted by error propagation and accumulation during each step; and the MAJ5 involved in producing S is inherently less accurate than the MAJ3.
- Both $C_{out}$ and S have higher accuracy when the input [ABC] = 000 or 111 than in the other cases. This is expected since the input states of all '0's and all '1's yield higher accuracy than these with mixed numbers of '0's and '1's for both MAJ3 and MAJ5.

The experimental results for the all-NAND design are shown in Fig. 5d. The same observations regarding accuracy vs. inputs as the MAJ+NOT design apply. However, it is clear that the accuracy of the all-NAND full adder, at 78.5%, is higher than that of the MAJ+NOT full adder, at 63.8%. This is likely due to the fact that 2-input NAND operations are inherently more accurate than MAJ3 and MAJ5 operations. This offsets the impact of the additional steps required in the all-NAND design. We note that the accuracy of all computation blocks will improve as the underlying MTJ technology evolves. Accordingly, the relative accuracy of the all-NAND versus the MAJ+NOT designs may change[66].

*IV. Modeling and Analysis of CRAM Logic Accuracy*

To understand the origin of errors, how they accumulate, and how they propagate, we performed numerical simulations of the full adder designs. These are based on probabilistic models of logic operations, implemented by Monte Carlo methods. Figures 5e and 5f show the simulation results for the MAJ+NOT and all-NAND design, respectively. In these, the accuracy of individual logic operations was set to match what was experimentally observed. The simulation results for the overall designs of the full adders correspond well to what was observed experimentally for these, which confirms the validity of the proposed probabilistic models (more details in Methods section and in Supplementary Note S6).

We note that beyond the inherent inaccuracy of logic operations, other factors such as device drift and device-to-device variation in MTJ devices will contribute to error in a CRAM. Specifically, drifts in temperature, external magnetic field, MTJ anisotropy, and MTJ resistance can lead to drift of the response curve, $<D_{out}>$. Most likely, any such drift will result in a reduction (increase) of accuracy (error rate). More discussion regarding device-to-device variation is provided Supplementary Note S7.

On the other hand, the accuracy of logic operations will significantly benefit from improvements in TMR ratio as the MTJ technology evolves. To project the future accuracy of CRAM operations, we employ various types of physical modeling informed by existing experimental results (more details are given in the Methods section and Supplementary Note S8).

Three sets of assumptions on the accuracies (or error rates) of NAND logic operations underlie the following studies.
- The 'experimental' assumptions are based on the best accuracy experimentally observed among the 9 NAND steps involved with the all-NAND 1-bit full adder. These are adjusted linearly to ensure that the error for inputs '01' and '10' equals that for input '11'. In reality, as supported by the experimental results shown in fig. 4a, such a condition can be reached by properly tuning the $V_{logic}$. Therefore, assuming the gate-level error rate is already optimized by tuning the $V_{logic}$, then the per-input-state NAND accuracies can be further simplified so that an error rate, $\delta$ ($0 \leq \delta \leq 1$), can be used to characterize the error, accuracy, and probabilistic truth table of NAND operations in a CRAM. The NAND accuracy is [1, 1-$\delta$, 1-$\delta$, 1-$\delta$], and the NAND probabilistic truth table is [1, 1-$\delta$, 1-$\delta$, $\delta$], both being a function of $\delta$. Through the above-mentioned modeling and calculations, the 'experimental' assumptions yield $\delta$ = 0.0076, which corresponds to a TMR ratio of approximately 109%, based on experiments.

- Two additional sets of assumptions, labelled as 'production', and 'improved' assume MTJ TMR ratios of 200%, and 300%, respectively. These two assumptions yield $\delta = 2.1 \times 10^{-4}$, and $\delta = 7.6 \times 10^{-6}$, respectively, based on modeling and calculations. The 'production' assumptions represent the current industry-level TMR ratios developed for STT-MRAM technologies. The 'improved' assumptions present reasonable expectations for near-future MTJ developments.

CRAM NAND error rates vs. TMR ratio with various logic voltage pulse widths are shown in Fig. 6a. Higher TMR ratios and faster logic speed – so shorter $V_{logic}$ pulse widths – lead to smaller error rates. Further details can be found in Supplementary Note S8 and in Supplementary Figure S5. Also included there is an analysis of error rates vs. effective TMR ratio, which is independent of the specific TMR modeling. Note that, for all subsequent results, we will use the NAND error rate at the assumed TMR ratios, with pulse widths of 1 ms. This is more conservative but is consistent with the pulse widths used in the experimental results reported above.

### V. Analysis of CRAM multi-bit adder, multiplier, and matrix multiplier

With these defined sets of assumptions, we provide projections of CRAM accuracy at a larger scale for meaningful applications. First we evaluate ripple-carry adders and array multipliers[72] operating on scalar operands, with up to 6 bits. To evaluate the results, we adopt the normalized error distance (NED) metric[73] to represent the error of these primitives, since it was shown to be more suitable for arithmetic primitives in presence of computational error. We will refer to the error for a given primitive as 'NED error'. We also define a complementary metric of 'NED accuracy' as the NED subtracted from 1 and then multiplied by 100%, to facilitate more intuitive visualization of the error values. While the 'experimental' assumptions with TMR ratio of 109% yield good overall accuracy for adders and multipliers, as the TMR ratio increases, the 'production' assumption with TMR ratio of 200%, and the 'improved' assumption with TMR ratio of 300%, yield significantly better or higher accuracies. Specifically, a 4-bit adder produces NED error of $2.8 \times 10^{-2}$, $8.6 \times 10^{-4}$, and $3.3 \times 10^{-5}$, or NED accuracy of 97.2%, 99.914%, and 99.9967%, for the 'experimental', 'production', and 'improved' assumptions, respectively. A 4-bit multiplier produces NED error of $5.5 \times 10^{-2}$, $1.8 \times 10^{-3}$, and $6.6 \times 10^{-5}$, or NED accuracy of 94.5%, 99.82%, and 99.9934%, for the three sets of assumptions, respectively. It is as expected that when comparing the adder to the multiplier, since the latter is more complex and involves more gates, its accuracy is generally lower than that of the adder. Similarly, when the bit width of the adder or multiplier increases, their accuracy decreases. Further details and results with bit width up to 6-bit are provided in the Methods section and in Supplementary Note S9.

Then, with these primitives, we evaluate dot-product operations, which form the basis of matrix multiplication. It is heavily employed in many applications in both conventional domains and machine intelligence. Dot products consist of element-wise multiplication of two unsigned integer vectors, followed by addition. We perform additions with binary trees to maintain smaller circuit depth. Figure 6b shows the NED error of a 4-bit 4×4 dot-product matrix multiplier with respect to various TMR ratio assumptions. Like the adders and multipliers, as the TMR ratio increases, the NED error decreases, or the NED accuracy improves. Specifically, a 4-bit 4×4 dot-product matrix multiplier produces NED error of 0.11, $3.4 \times 10^{-3}$, and $1.2 \times 10^{-4}$, or NED accuracy of 89%, 99.66%, and 99.988%, for the 'experimental', 'production', and 'improved' assumptions, respectively. Also, when comparing the different input sizes (e.g., 1×1 to 4×4), as expected, the NED error is larger for larger input sizes due to the increased number of gates involved. Further details and results with bit width up to 5-bit are provided in the Methods section and in Supplementary Note S9.

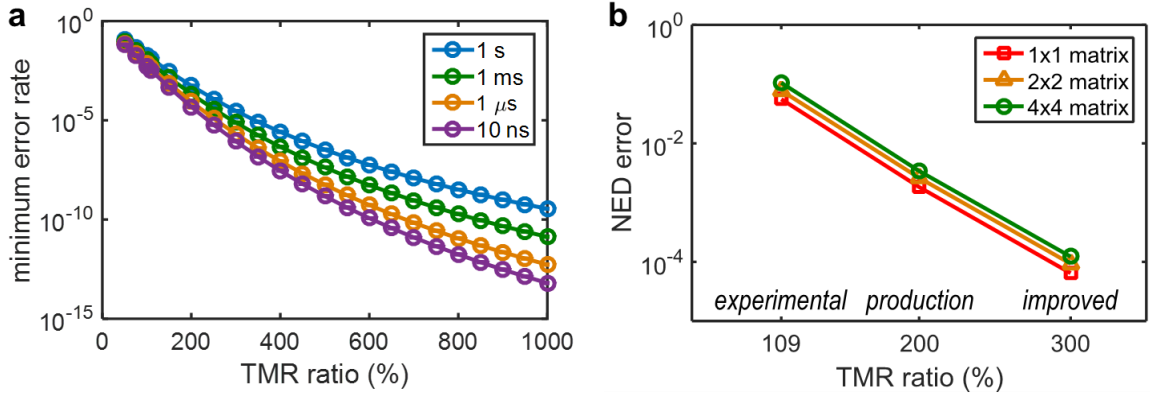

**Fig. 6 | CRAM accuracy projections. a** NAND gate minimum error rate vs. MTJ TMR ratio with various $V_{logic}$ pulse widths. For a given TMR ratio, the error rate is a function of $V_{logic}$. So, the 'minimum error rate' represents the minimum error rate achievable with an appropriate $V_{logic}$ value. All subsequent studies use the error rates observed with 1 ms pulse widths (to be consistent with the earlier experimental studies) at assumed TMR ratios. **b** The NED error of 4-bit dot-product matrix multiplier vs. TMR ratio. TMR ratios of 109%, 200%, and 300% are adopted for the 'experimental', 'production, and 'improved' assumptions, respectively. The size of the input matrix is indicated in the legend of the plot.

## Discussions

To summarize the experimental work, an MTJ-based 1×7 CRAM array hardware was experimentally demonstrated and systematically evaluated. The basic memory write and read operations of CRAM were achieved with high reliability. The study on CRAM logic operations began with 2-input logic operations. It was found that a 2-input NAND operation could be performed with accuracy as high as 99.4%. As the number of input cells was increased, for example, for 3-input MAJ3 and 5-input MAJ5 operations, the accuracy decreased to 86.5% and 75%, respectively. The decrease was attributed to having too many levels corresponding to the input states crowding a limited operating margin. Next, two versions of a 1-bit full adder were experimentally demonstrated using the 1×7 CRAM array: an all-NAND version and a MAJ+NOT version. The all-NAND design achieved an accuracy of 78.5% while the seemingly simpler MAJ+NOT, which involves 3- and 5-input MAJ operations, only achieved an accuracy of 63.8%. Note that although each type of logic operation achieves optimal accuracy performance with a specific voltage value, the value is expected to only need to be static or constant. Therefore, only a finite number of power rails is needed to accommodate the logic operations of the CRAM array. Also, if the multiple logic pulse duration is allowed by a peripheral design, it is possible to operate the CRAM array with a single set of power rails for both memory write and logic operations.

A probabilistic model was proposed that accounts for the origin of errors, their propagation, and their accumulation during a multi-step CRAM operation. The model was shown to be effective when matched with the experimental results for the 1-bit full adder. The working principles of this model were adopted for the rest of the studies.

A suite of MTJ device circuit models were fitted to the existing experimental data and used to project CRAM NAND gate-level accuracy in the form of error rates. The gate-level error rates were shown to be $7.6\times10^{-6}$, with reasonable expectations of TMR ratio improvement as MTJ technology develops. Other device properties, such as the switching probability transfer curve, could also significantly affect the CRAM gate-level error rate. This calls for improvements or new discoveries of the physical mechanisms for memory read-out, and memory write. Error is an inherent property of any physical hardware, including CMOS logic components, which are commonly perceived as deterministic. As the development of CRAM proceeds, the gate-level error rate of CRAM will be further reduced over time. For now, while the error rate of CRAM is still higher compared to that of CMOS logic circuits, CRAM is naturally more suitable for applications that require less precision but can still benefit from the true-in-memory computing features and advantages of CRAM, instead of those that require high precision and determinism. Additionally, logic operations with many inputs, such as majority, may be desirable in certain scenarios. And yet, these were shown to have lower accuracy than 2-input operations. So, a tradeoff might exist.

Lastly, building on the experimental demonstration and evaluation of the 1-bit full adder designs, simulation and analysis was performed for larger functional circuits: scalar addition and multiplication up to 6 bits and matrix multiplication up to 5 bits with input size up to 4×4. These are essential building blocks for many conventional and machine intelligence applications. The parameters for the simulations were experimentally measured values as well as reasonable projections for future MTJ technology. The results show promising accuracy performance of CRAM at a functional building block level. Furthermore, as device technologies progress, improved performance or new switching mechanisms could further reduce the gate-level error rate of CRAM. Error correction techniques could also be employed to suppress CRAM gate errors.

In summary, this work serves as the first step in experimentally demonstrating the viability, feasibility, and realistic properties of MTJ-based CRAM hardware. Through modeling and simulation, it also lays out the foundation for a coherent view of CRAM, from the device physics level up to the application level. Prior work had established the potential of CRAM through numerical simulation only. Accordingly, there had been considerable interest in the unique features, speed, power, and energy benefits of the technology. This study puts the earlier work on a firm experimental footing, providing application-critical metrics of gate-level accuracy or error rate and linking it to the application accuracy. It paves the way for future work on large scale applications, in conventional domains as well as new ones emerging in machine intelligence. It also indicates the possibility of making competitive large-scale CMOS-integrated CRAM hardware.

## Methods

*I. MTJ fabrication and preparation*

The MTJ thin film stacks were grown by magnetron sputtering in a 12-source deposition system with the base pressure of $5\times10^{-9}$ Torr. The MgO barrier was fabricated by RF sputtering while all the metallic layers were fabricated by DC sputtering. The stack structure is Si/SiO$_2$/Ta(3)/Ru(6)/Ta(4)/Mo(1.2)/Co$_{20}$Fe$_{60}$B$_{20}$(1)/MgO(0.9)/Co$_{20}$Fe$_{60}$B$_{20}$(1.4)/Mo(1.9)/Ta(5)/Ru(7) where numbers in brackets indicate thickness of the layer in nm. The stack was then annealed at 300 °C for 20 minutes in a rapid thermal annealing system under Ar atmosphere (more information on the MTJ stack fabrication can be found in refs.[74,75]).

The MTJ stacks were fabricated using three rounds of lithography similar to those described in ref. [76]. First, the bottom contacts were defined using photolithography followed by Ar+ ion milling etching. Then, the MTJ pillars were patterned into 120-nm circular nano-pillars via E-beam lithography and etched through Ar+ ion milling. After etching, SiO$_2$ was deposited via plasma enhanced chemical vapor deposition (PECVD) to protect the nano-pillars. Finally, the top contacts were defined using photolithography and the metallic electrodes of Ti (10 nm)/Au (100 nm) were deposited using electron beam evaporation.

The die of MTJ array was diced into smaller pieces with each piece containing approximately 10 MTJ devices. Each of the small pieces was mounted on a cartridge board and up to 8 MTJ devices were wire-bonded to the electrode of the cartridge board. Seven cartridge boards were inserted onto the connection board providing MTJs to the CRAM. The MTJ in each CRAM cell is elected among up to 8 MTJs on the corresponding cartridge board. In total, seven MTJs are elected from up to 56 MTJs. This method allows the user to find a collection of seven MTJs with minimum device-device variations.

*II. CRAM experiment*

Individual bias magnetic field was implemented for each of seven MTJs on the connection board by positioning a permanent magnet at a certain distance to the MTJ devices. Bias magnetic field was used to compensate intrinsic magnetic exchange bias and stray field in MTJ devices and to restore the balance between P and AP state. Additionally, slight rotation of bias field in the device plane was used to effectively adjust the switching voltage of each MTJ. More details can be found in Supplementary Note S2.

The connection board with seven MTJs was connected to the main board. On the main board, necessary active and passive electronic components were populated on the custom-designed PCB. The CRAM demo hardware circuit implemented a 1×7 CRAM array with a modified architecture to emphasize logic operations while compromising on memory operations bandwidth for simplicity. It was modified from the full-fledged 2T1M[40] architecture. It was equivalent to a 2T1M CRAM in logic mode, but it only had serial access to all cells for memory read and write operations (more details in Supplementary Note S1). The hardware was powered by a battery and communicated with the controller PC wirelessly via Bluetooth®. In this way, the entire hardware was electrically isolated from the environment so that the risk of ESD to these sensitive MTJs was minimized.

The experiment control software running on a PC was implemented by National Instruments' LabVIEW™. It was responsible for real-time measurements and real-time control of the experiments, as well as necessary visualizations. Certain results were further analyzed post-experiment.

*III. CRAM modeling and simulations*

The simulation studies of the accuracy or the error origination, accumulation, and propagation began with a simple probabilistic model of each NAND logic operation. A probabilistic truth table was used to describe the expected statistical average of the output logical state. Then the 1-bit full adder designs or operations were simulated by the Monte Carlo method and with assumed probabilistic truth tables for each of the logic steps (see Supplementary Note S6).

The experiment-based physics modeling and calculations for obtaining the projected CRAM logic operation accuracies began with an MTJ resistance-voltage model[77] which was fit to the experimental data of TMR vs. bias voltage. The coefficients of this model were scaled accordingly to model projected TMR ratios higher than that of the experimental observation. Then a thermal activation model[78,79] of MTJ switching probability was fit to experimental data and was used to calculate the switching probability of the output MTJ cell under various bias voltage. Finally, the average of output state, <$D_{out}$>, could be calculated under various $V_{logic}$ and the optimal NAND accuracies could be obtained in similar way as discussed with Fig. 4 (more details in Supplementary Note S8).

Further simulation studies of a ripple-carry adder, a systolic multiplier, and simulated the dot-product operation of a matrix multiplication for various number of bits as well as matrix sizes were carried out with the same methods. More details can be found in Supplementary Note S9.

## Acknowledgements

This work was supported in part by the Defense Advanced Research Projects Agency (DARPA) via No. HR001117S0056-FP-042 "Advanced MTJs for computation in and near random access memory" and by the National Institute of Standards and Technology. This work was supported in part by NSF SPX grant no. 1725420. The authors also thank Cisco Inc. for the support. Portions of this work were conducted in the Minnesota Nano Center, which was supported by the National Science Foundation through the National Nanotechnology Coordinated Infrastructure (NNCI) under Award No. ECCS-2025124. The authors acknowledge the Minnesota Supercomputing Institute (MSI, URL: http://www.msi.umn.edu) at the University of Minnesota for providing resources that contributed to the research results reported within this paper. The authors thank Prof. Marc Riedel and Prof. John Sartori from the Department of Electrical and Computer Engineering at University of Minnesota for proofreading the manuscript. Yang Lv, Brandon Zink, and Hüsrev Cılasun are CISCO Fellow.

## Author Contributions

J.-P.W. conceived the CRAM research and coordinated the entire project. Y.L. and J.-P.W. designed the experiments. Y.L. and R.P.B. designed and developed the demonstration hardware and software. P.K., A.H. and W.W. grew part of the perpendicular MTJ stacks. B.R.Z. fabricated the MTJ nano devices. Y.L. conducted the CRAM demonstration experiments and analyzed the results. Y.L. studied the probabilistic model of CRAM operations and conducted simulations of 1-bit full adder. Y.L., B.R.Z., and R.P.B developed the device physics modeling of CRAM logic operations and gate-level error rate and conducted related calculations. H.C., S.R., Z.C., and U.K. carried out the simulation studies of multi-bit adder, multiplier, and matrix multiplication. S.S. participated in discussions of modeling and simulation. All authors reviewed and discussed the results. Y.L. and J.-P.W. wrote the draft manuscript. All authors contributed to the completion of the manuscript.

## Competing Interest Statement

All authors declare no competing interests.

## Data availability

The authors declare that the data supporting the findings of this study are available within the paper, its supplementary information files.